\documentclass[12pt]{article}
\usepackage{amssymb}
\usepackage{graphics,mathrsfs}
\usepackage{graphicx,epstopdf,float}
\usepackage{amsmath,bm}
\usepackage{color}
\usepackage{placeins}
\usepackage{enumerate}
\usepackage{enumitem}
\usepackage{varioref}
\usepackage{indentfirst}
\usepackage{setspace}
\usepackage[title]{appendix}
\usepackage{amsfonts}
\usepackage{amsmath}
\usepackage{booktabs,tabularx}
\usepackage{subcaption}
\usepackage{caption}
\usepackage{lscape}
\usepackage{lineno}
\usepackage{multirow}  
\usepackage[authoryear]{natbib}
\setcitestyle{authoryear,open={(},close={)}}
\makeatletter \oddsidemargin  -.1in \evensidemargin -.1in
\begin{document}
	\newcommand{\bea}{\begin{eqnarray}}
		\newcommand{\eea}{\end{eqnarray}}
	\newcommand{\nn}{\nonumber}
	\newcommand{\bee}{\begin{eqnarray*}}
		\newcommand{\eee}{\end{eqnarray*}}
	\newcommand{\lb}{\label}
	\newcommand{\nii}{\noindent}
	\newcommand{\ii}{\indent}
	\newtheorem{thm}{Theorem}[section]
	\newtheorem{example}{Example}[section]
	\newtheorem{cor}{Corollary}[section]
	\newtheorem{definition}{Definition}[section]
	\newtheorem{lem}{Lemma}[section]
	\newtheorem{rem}{Remark}[section]
	\newtheorem{proposition}{Proposition}[section]
	\renewcommand{\theequation}{\thesection.\arabic{equation}}
	\bibpunct[, ]{(}{)}{;}{a}{,}{,}
	\renewcommand\bibfont{\fontsize{10}{12}\selectfont}
	\setlength{\bibsep}{0.0pt}
	
	\date{}
	\title{\bf The Topp-Leone XLindley Distribution: Properties, Estimation and Applications to Lifetime Data}
	
	\author{\large { \bf Saileshwari M$^1$\thanks{Email address: saileshwarimohan@gmail.com},~Rajesh Moharana$^1$\thanks{Email address (Corresponding author): rajeshmoharana31@gmail.com,~rajesh.moharana@vit.ac.in} and ~Kousik Maiti $^2$\thanks {Email address:  kousikulu@gmail.com, kousik.maiti@hithaldia.ac.in}  }
		\\{\it \small $^1$Department of Mathematics, School of Advanced Sciences, Vellore Institute of Technology,}\\{\it \small Vellore-632014, Tamil Nadu, India}
		\\{\it \small $^2$School of Applied Science $\&$ Humanities, Haldia Institute of Technology, Haldia-721657, India}}
\maketitle
\begin{center}
	{\large \bf Abstract}
\end{center}
In this article, we use a family of distributions developed by Topp-Leone to create a novel lifetime two-parameter distribution. The Topp-Leone XLindley distribution is the name used for it. We study several types of statistical and mathematical characteristics of this distribution, such as reliability functions, moments,  moment-generating function, quantile function, and the R\'{e}nyi entropy. For the purpose of estimating parameters under the proposed distribution model, this simulation study is conducted to evaluate the performance of the maximum likelihood estimates for the parameters of the Topp-Leone XLindley distribution. A comprehensive simulation investigation is carried out to evaluate the performance of the proposed methods. Furthermore, a practical data set of bank customer waiting times has been investigated for the purpose of illustration.
\section{Introduction}
Modelling lifetime data is of great importance in various fields such as reliability engineering, survival analysis, and operations research. Classical distributions such as the exponential (see \cite{balakrishnan1996exponential}) and Weibull are widely used because of their mathematical simplicity. However, they often fail to represent complex real-world behavior. In particular, they struggle to capture skewness, heavy tails, and non-monotonic hazard rates. To address these limitations, the Lindley distribution proposed by \cite{ghitany2008lindley} has been widely adopted as a flexible alternative to the exponential distribution. Despite its usefulness, it remains inadequate for modeling complex and heterogeneous data patterns, necessitates the development of more generalized forms. To further advance modeling performance, \cite{chouia2021xlindley} developed the X-Lindley (XL) distribution by combining the exponential and Lindley distributions, offering enhanced flexibility while retaining the same number of parameters. A random variable, $X$ is said to be distributed according to the XL distribution, if the cumulative distribution function (CDF) and probability density function (PDF) of $X$ are correspondingly given by
\begin{eqnarray}\label{xcdf}
	F_{XL}(x|\theta)&=&1-\biggl(1+\frac{\theta x}{(1+\theta)^2}\biggr) \exp\{-\theta x\},~ x>0,~\theta>0;
\end{eqnarray}
\begin{eqnarray}\label{xpdf}
\mbox{and}~~	f_{XL}(x|\theta)&=& \frac{\theta^2(2+x+\theta)\exp\{-\theta x\}}{(1+\theta)^2},~x>0,~\theta>0.
\end{eqnarray}
The XL distribution has shown improved performance over the exponential and Lindley distributions, and other competing models in various lifetime data applications. However, its structure still imposes limitations when it comes to modeling more complex hazard rate behaviors. In recent years, the XL distribution has inspired several generalizations to accommodate increasingly complex data structures. For example, \cite{ahsan2022poisson} formulated the Poisson XLindley distribution for handling discrete count data. Similarly, \cite{meriem2022retracted} developed the Power XLindley model within a fuzzy reliability context and proposed nine parameter estimation techniques. \cite{metiri2022characterisation} characterized the XL distribution using truncated moments, while \cite{alotaibi2022computational} focused on parameter estimation under adaptive Type-II progressive hybrid censoring. In another extensions, \cite{khodja2022truncated} explored a truncated XLindley distribution and \cite{beghriche2023inverse} introduced the Inverse XLindley distribution.

A widely used strategy for increasing distributional flexibility is the generator-based approach, in which a generator function is applied to a baseline distribution to construct a more flexible model through the introduction of additional parameters. Among these methods, the Topp–Leone generator, originally proposed by \cite{topp1955} as a J-shaped distribution for modelling data on the unit interval, has received considerable attention for its effectiveness in enriching existing distributions. Over time, it has been successfully employed to generalize numerous models, substantially improving their ability to accommodate diverse data characteristics. For instance, several studies have examined its theoretical properties in detail. \cite{ghitany2005} investigated its reliability characteristics, \cite{genc2012} analyzed its order statistics, while \cite{ahmad2015} and \cite{david2017} proposed exponentiated forms and improved estimation techniques. Building on this foundation, \cite{alshomrani2016} introduced the Topp-Leone generated family, which has since inspired several extensions such as the Topp-Leone Gumbel (\cite{winai2016}) and Topp-Leone Burr XII (\cite{hesham2017}) distributions, further demonstrating its modeling potential.

Although the Topp–Leone generator has been extensively employed to extend and enrich several classical lifetime distributions, its application to the XLindley distribution has not yet been explored in the existing literature. This gap motivates the present study, which proposes a novel two-parameter lifetime distribution model, referred to as the Topp-Leone XLindley (TLXL) distribution. This novel distribution is obtained by applying the Topp-Leone generator to the XLindley distribution. The resulting TLXL model introduces a shape parameter in addition to the scale parameter. This makes the distribution more adaptable and able to capture the various lifetime data behaviors. By compounding the Topp-Leone distribution with the XLindley distribution, the proposed distribution model is more adaptable and can capture various forms of the hazard rate functions, such as increasing, decreasing, and bathtub-shaped hazard rate functions. The objectives of the present study are presented as follows:

\begin{itemize}
	\item[$(i).$] To construct the TLXL distribution, and investigate its key statistical and reliability properties.
	\item[$(ii).$] To estimate the model parameters using the maximum likelihood method and assess the performance 
    of the estimators through simulation studies.
	\item[$(iii).$] To evaluate the practical usefulness of the TLXL distribution by analyzing real lifetime data and 
comparing its fitting performance with existing competing models.
\end{itemize}
The remainder of the paper is organized as follows. Section $2$ presents the construction of the TLXL distribution and examines its limiting behavior. Section $3$ discusses various statistical properties and reliability characteristics of the proposed model. In Section $4$, the maximum likelihood estimation procedure is developed, along with a simulation study to assess the performance of the estimators. Section $5$ illustrates the applicability of the TLXL distribution using the bank customer waiting time datasets. Finally, Section $6$ concludes the paper and outlines possible directions for future research.
	
\section{The Topp-Leone XLindley Distribution }\setcounter{equation}{0}
This section introduces the Topp-Leone XLindley (TLXL) distribution by presenting its probability density function (PDF) and cumulative distribution function (CDF), followed by graphical illustrations and an examination of its limiting behavior.
\subsection{Construction of the TLXL Model}
In this section, we propose a new two-parameter extension of the XLindley distribution by employing the Topp--Leone generator. Specifically, the XLindley distribution is taken as the baseline model within the Topp--Leone generated (TL-G) family of distributions introduced by \cite{al2016toppa}, leading to the formulation of the TLXL distribution with parameters $\theta$ and $\alpha$. If $X$ is a continuous random variable with CDF $G(x)$ and PDF $g(x)$, then the CDF and PDF of the TL–G family of distributions are given by
	\begin{equation}\label{topcdf}
		G_{TL-G}(x|\alpha) = [2F(x) -( F(x))^2]^{\alpha} = \left[1 - (\bar{F}(x))^2\right]^{\alpha}, ~0 < x < 1, \alpha > 0;
	\end{equation}
	\begin{equation}\label{toppdf}
		\mbox{and}~~g_{TL-G}(x|\alpha) = 2\alpha f(x) \bar{F}(x) \left[1 - (\bar{F}(x))^2\right]^{\alpha - 1}, ~ 0 < x < 1, 
\alpha > 0;
	\end{equation}
	where $f(x)$, $F(x)$ and $\bar{F}(x) = 1 - F(x)$  represent the PDF, CDF and survival function of the baseline distribution, respectively. Now, by substituting the CDF and PDF of the XLindley distribution, given respectively in (\ref{xcdf}) and (\ref{xpdf}), into (\ref{topcdf}) and (\ref{toppdf}), we obtain the CDF and PDF of the TLXL distribution as follows:
		\begin{eqnarray}\label{cdf}
		G_{TLXL}(x|\alpha,\theta)&=&\left(1-\left(\phi(\theta, x)\exp\{-\theta x\}\right)^2\right)^\alpha
	\end{eqnarray}
	\begin{eqnarray}\label{pdf}
		\mbox{and}~~ g_{TLXL}(x|\alpha,\theta)= \frac{2\alpha\theta^2 \phi(\theta, x)\exp\{-2\theta x\} (2+x+\theta)}{(1+\theta)^2}\left(1-\left(\phi(\theta, x)\exp\{-\theta x\}\right)^2\right)^{\alpha-1} 
	\end{eqnarray}
where $\phi(\theta, x)= \frac{(1+\theta)^2+\theta x}{(1+\theta)^2}, 0<x<1$, and $\theta,\alpha>0$. Note that, in the TLXL distribution, $\alpha$ serves as the shape parameter, while $\theta$ is the scale parameter. Now, we visualize the PDF and CDF of the TLXL distribution for various parameter values. These plots clearly illustrate the ability of the TLXL distribution in capturing different distributional shapes, skewness patterns, and tail behaviors commonly observed in lifetime and reliability data.

	\begin{figure}[H]
		\centering
		\subcaptionbox{}[3in]{\includegraphics[height=2.8in, width = 3in]{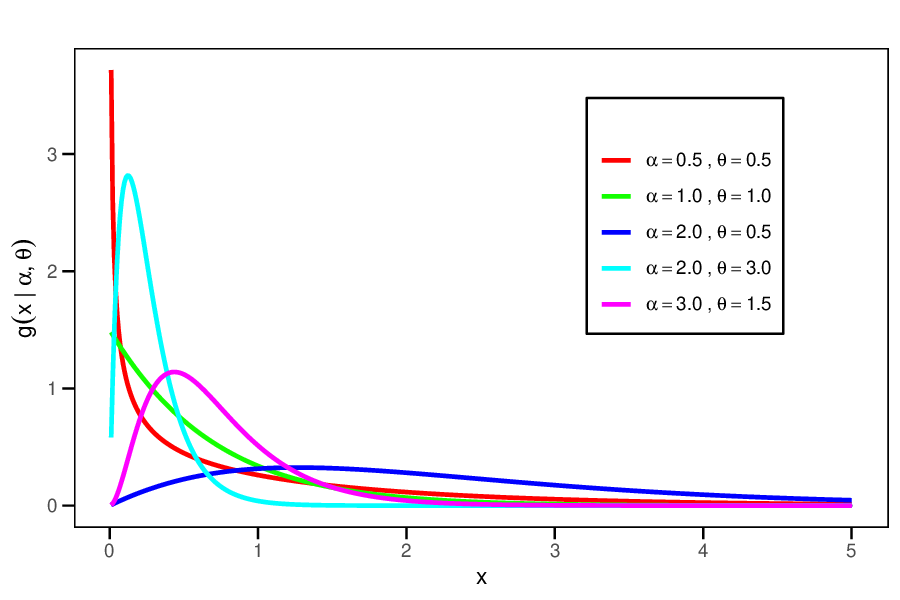}}
		\subcaptionbox{}[3in]{\includegraphics[height=2.8in, width = 3in]{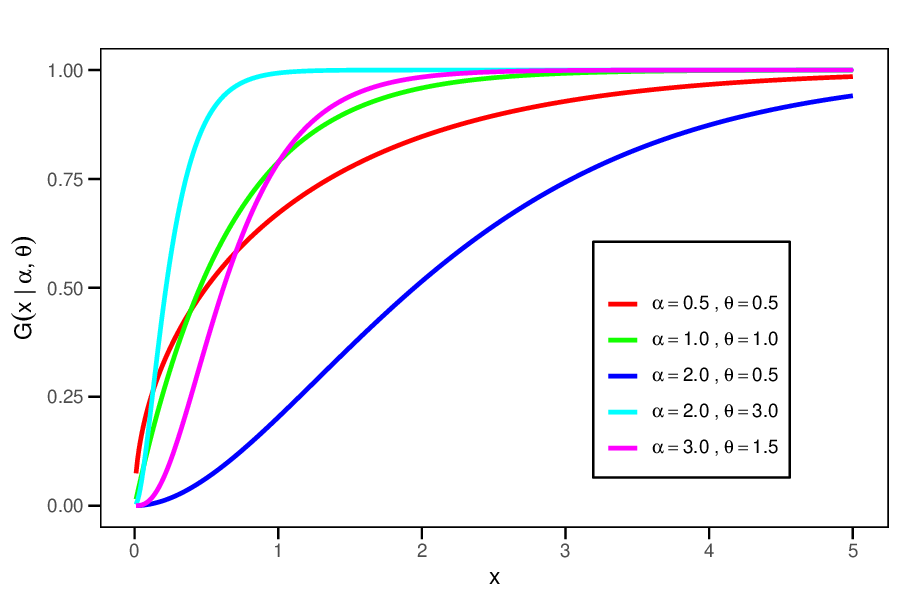}}
		\caption{$(a)$ represents the PDF plots of the TLXL distribution for various combinations of $\theta$ and $\alpha$, 
        and $(b)$ represents the CDF plots of the TLXL distribution for different values of $\theta$ and $\alpha$.}
		\label{pdfplot}
	\end{figure}
From Figure~\ref{pdfplot}$(a)$, it is observed that the PDF of the TLXL distribution exhibits flexible behavior depending on its parameters. When the shape parameter $0<\alpha < 1$, the distribution is monotonically decreasing. For $\alpha > 1$, it becomes uni-modal with a distinct peak. The scale parameter $\theta$ influences the spread. The larger values of $\theta$ yield sharper peaks and lighter tails, whereas smaller values result in flatter shapes with heavier tails. Figure~\ref{pdfplot}$(b)$ shows that the CDF of the TLXL distribution displays a variety of shapes based on the parameter values. As the shape parameter $\alpha$ increases, the distribution reaches its upper bound more quickly, indicating faster accumulation of probability mass. Similarly, higher values of the scale parameter $\theta$ steepen the curve, suggesting quicker progression toward the maximum value. This flexibility makes the TLXL distribution a valuable tool for modeling skewed or peaked lifetime data, particularly in reliability and survival analysis.

Next, we express the PDF and CDF of the TLXL distribution as infinite linear combinations of exponential functions derived via the Binomial series expansion. To this end, consider the Binomial series expansion $(1 + x)^a = \sum_{k=0}^{\infty} \binom{a}{k} x^k$ for $|x| < 1$. Thus, the CDF and PDF of the TLXL distribution, correspondingly given in (\ref{pdf}) and (2.4), can be expressed as:
\begin{eqnarray}
	G_{TLXL}(x|\alpha,\theta) &=& \sum_{l=0}^{\infty}\sum_{m=0}^{2l}  x^m e^{-2l\theta x} \Psi_{l,m} \nonumber\\
\mbox{and}~~	g_{TLXL}(x|\alpha,\theta) &=& \frac{2\alpha \theta^2}{(1+\theta)^2} \sum_{l=0}^{\infty}\sum_{m=0}^{2l+1}(2+\theta+x)x^{m} e^{-2\theta(l+1)x} \Lambda_{l,m} \nonumber
\end{eqnarray}
where $\Psi_{l,m}=\binom{\alpha}{l} \binom{2l}{m} (-1)^l \left(\frac{\theta}{(1+\theta)^2}\right)^m$ and $\Lambda_{l,m}= \binom{\alpha-1}{l} \binom{2l+1}{m}(-1)^l \left(\frac{\theta}{(1+\theta)^2}\right)^m$.

\subsection{Properties of the TLXL Distribution \setcounter{equation}{0}}	
In this subsection, we present the reliability and statistical properties of the TLXL distribution.

First, we obtain various reliability measures associated with the TLXL distribution. Specifically, we derive expressions for the survival function $\left(\bar{G}_{TLXL}(x|\alpha, \theta)\right)$, hazard rate function $\left(h_{TLXL}(x|\alpha, \theta)\right)$, reverse hazard rate function $\left(rhr_{TLXL}(x|\alpha, \theta)\right)$ and cumulative hazard rate function $\left(H_{TLXL}(x|\alpha, \theta) \right)$, which are correspondingly given below:
\begin{eqnarray}
	\bar{G}_{TLXL}(x|\alpha,\theta) &=& 1-G_{TLXL}(x|\alpha, \theta)
	= 1-\left(1-\left(\phi(\theta,x) \exp\{-\theta x\}\right)^2\right)^\alpha, \label{sur}\\
	 h_{TLXL}(x|\alpha,\theta) &=& \frac{g_{TLXL}(x|\alpha, \theta)}{\bar{G}_{TLXL}(x|\alpha, \theta)} \nonumber\\
	 &=&\frac{2\alpha\theta^2 (2+x+\theta)\phi(\theta,x) \exp\{-2\theta x\}\left(1-\left(\phi(\theta,x) \exp\{-\theta x\}\right)^2\right)^{\alpha-1}}{(1+\theta)^2\left(1-\left(1-\left(\phi(\theta,x) \exp\{-\theta x\}\right)^2\right)^\alpha\right)}, \label{ha}\nonumber\\
	 &~&\\
	 rhr_{TLXL}(x|\alpha,\theta) &=& \frac{g_{TLXL}(x|\alpha, \theta)}{G_{TLXL}(x|\alpha, \theta)} =\frac{2\alpha\theta^2(2+x+\theta)\phi(\theta,x) \exp\{-2\theta x\}}{(1+\theta)^2\left(1-\left(\phi(\theta,x) \exp\{-\theta x\}\right)^2\right)}, \label{rha}\\
	H_{TLXL}(x|\alpha,\theta) &=& -\ln(\bar{F}_{TLXL}(x|\alpha, \theta)) = -\ln\left(1-\left(1-\left(\phi(\theta,x) 
 \exp\{-\theta x\}\right)^2\right)^\alpha\right) \label{cha}
\end{eqnarray}
where $\phi(\theta, x)= \frac{(1+\theta)^2+\theta x}{(1+\theta)^2}, 0<x<1$, and $\theta,\alpha>0$. In the following, we present graphical illustrations of the hazard rate function to demonstrate its behavior under different parameter settings.
\begin{figure}[H]
	\centering
	\includegraphics[width=0.75\textwidth]{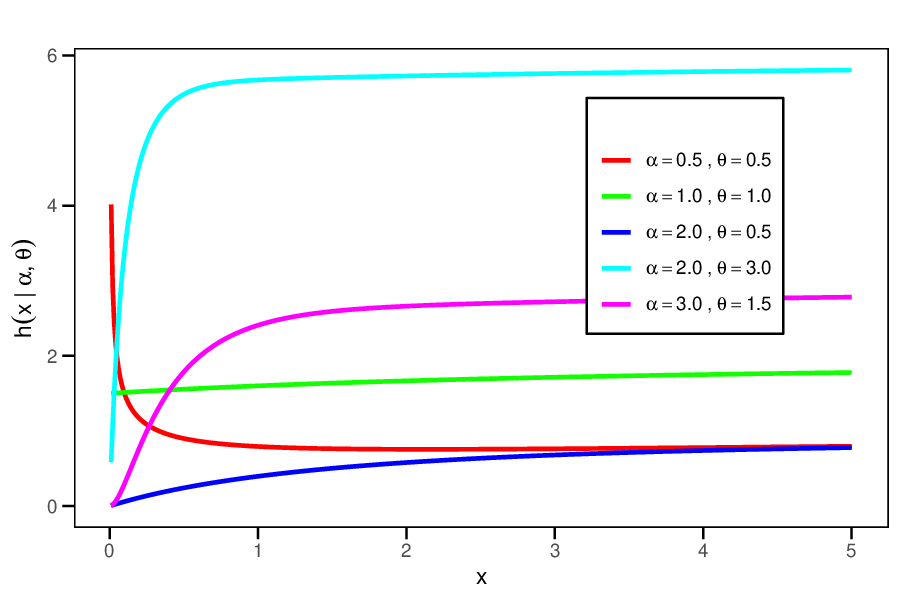}
	\caption{Hazard rate plot of the TLXL distribution for various values of $\theta$ and $\alpha$.}
	\label{haplot}
\end{figure}
As seen in Figure~\ref{haplot}, the hazard rate of the TLXL distribution exhibits various shapes depending on the parameters $\alpha$ and $\theta$. For $\alpha < 1$, the hazard rate function tends to be decreasing, indicating early failures. When $\alpha > 1$ and $\theta$ is moderate or small, the hazard rate becomes increasing, representing aging systems. A bathtub-shaped hazard, capturing both early and late failures, typically arises when $\alpha \approx 1$ and $\theta$ is moderate.

Next, the moments and associated measures of the TLXL distribution are presented. The $r$-{th} moment of a continuous random variable $X$ which satisfy the TLXL distribution can be defined as:
\begin{eqnarray*}
	\mu_r^{'} = E(X^r)=\int_{0}^{1} x^r g_{TLXL}(x|\alpha,\theta) dx.
\end{eqnarray*}
Now, substituting the PDF of the TLXL distribution provided in (2.2), we will get the $r$-{th} moment as:
\begin{eqnarray}\label{eq3.5}
	\mu_r^{'} = \frac{2\alpha \theta^2}{(1+\theta)^2} \sum_{l=0}^{\infty} \sum_{m=0}^{2l+1} (-1)^l \binom{\alpha-1}{l} \binom{2l+1}{m} \bigg(\frac{\theta}{(1+\theta)^2}\bigg)^m 
	\int_{0}^{1} \frac{x^{m+r} (2+\theta+x)}{\exp\{2\theta(l+1)x\}} dx.
\end{eqnarray}
Further, using the Taylor series expansion of the exponential function, i.e.,
\[
\exp\{-tx\}=\sum_{n=0}^{\infty} (-1)^{n}\frac{t^n x^n}{n!},
\]
into (\ref{eq3.5}), we have
\begin{eqnarray}\label{mom}
	\mu_{r}^{'} &=& \frac{2\alpha \theta^2}{(1+\theta)^2} \sum_{l=0}^{\infty} \sum_{m=0}^{2l+1} \sum_{n=0}^{\infty} \frac{(-1)^{l+n}(2\theta(l+1))^n}{n!} \binom{\alpha-1}{l} \binom{2l+1}{m} \bigg(\frac{\theta}{(1+\theta)^2}\bigg)^m  \nonumber\\
	&~& \times \int_{0}^{1} x^{m+r+n} (2+\theta+x) dx.
\end{eqnarray}

Now, we rewrite the integral term present in (\ref{mom}) in terms of Beta function, defined as $\beta(n, m) = \int_0^1 t^{n-1}(1 - t)^{m-1} dt$ (see \cite{gradshteyn2014table}), which leads to the following expression for the moments:
\begin{eqnarray}
\mu_r^{'}
 &=& \frac{2\alpha \theta^2}{(1+\theta)^2} \sum_{l=0}^{\infty} \sum_{m=0}^{2l+1} \sum_{n=0}^{\infty} \frac{(-1)^{l+n}(2\theta(l+1))^n}{n!} \binom{\alpha-1}{l} \binom{2l+1}{m} \bigg(\frac{\theta}{(1+\theta)^2}\bigg)^m \nonumber \\&~&\times ((2+\theta) \beta(m+r+n+1,1)+\beta(m+r+n+2,1)). \label{mf}
\end{eqnarray}
The mean $(\mu)$, variance $(\sigma^2)$, coefficient of variance $(CV)$, skewness $(\gamma_1)$, and kurtosis $(\gamma_2)$ of the TLXL distribution are obtained from the $r$-th raw moment by setting $r = 1, 2, 3, 4$. These measures are expressed as:
\begin{eqnarray*}
	\mu &=& \mu'_1, \quad
	  \sigma^2 = \mu'_2 - (\mu'_1)^2, \quad
	CV = \sigma / \mu,
\end{eqnarray*}
\begin{eqnarray*}
	\gamma_1 &=& \frac{\mu'_3 - 3\mu'_2\mu'_1 + 2(\mu'_1)^3}{(\sigma^2)^{\frac{3}{2}}}, \quad
	\gamma_2 = \frac{\mu'_4 - 4\mu'_3\mu'_1 + 6\mu'_2(\mu'_1)^2 - 3(\mu'_1)^4}{(\sigma^2)^2}.
\end{eqnarray*}
Owing to the analytical complexity of the closed-form expressions of these measures, we report the numerical values of the first four moments and their associated measures of the TLXL distribution for various choices of $\theta$ and $\alpha$.
\begin{table}[H]
	\centering
	\caption{Numerical values of first four moments and its associated measures of the TLXL distribution for different values of $\theta$ and $\alpha$.}
	\begin{tabular}{cc|cccc|cc|cc}
		\hline
		$\theta$ & $\alpha$ & $\mu'_1$ & $\mu'_2$ & $\mu'_3$ & $\mu'_4$ & $\sigma^2$ & $CV$ & $\gamma_1$ &  $\gamma_2$ \\
		\hline
		0.7 & 0.5 & 0.2193 & 0.1211 & 0.0825 & 0.0623 & 0.0730 & 1.2320 & 1.2121 & 3.3523 \\
		& 1.0 & 0.2656 & 0.1634 & 0.1167 & 0.0903 & 0.0928 & 1.1471 & 0.8472 & 2.3893 \\
		& 1.5 & 0.2556 & 0.1691 & 0.1252 & 0.0990 & 0.1037 & 1.2599 & 0.8660 & 2.2743 \\
		& 2.0 & 0.2268 & 0.1581 & 0.1204 & 0.0969 & 0.1066 & 1.4399 & 1.0396 & 2.5082 \\
		
		\midrule
		1.0 & 0.5 & 0.2164 & 0.1118 & 0.0733 & 0.0540 & 0.0650 & 1.1781 & 1.2661 & 3.6319 \\
		& 1.0 & 0.3002 & 0.1731 & 0.1188 & 0.0896 & 0.0830 & 0.9595 & 0.7135 & 2.3475 \\
		& 1.5 & 0.3297 & 0.2048 & 0.1457 & 0.1121 & 0.0961 & 0.9404 & 0.4970 & 1.9593 \\
		& 2.0 & 0.3331 & 0.2186 & 0.1600 & 0.1253 & 0.1076 & 0.9850 & 0.4398 & 1.7809 \\
		
		\midrule
		1.2 & 0.5 & 0.2062 & 0.1015 & 0.0647 & 0.0467 & 0.0590 & 1.1781 & 1.3543 & 3.9826 \\
		& 1.0 & 0.3010 & 0.1655 & 0.1103 & 0.0816 & 0.0749 & 0.9096 & 0.7521 & 2.5139 \\
		& 1.5 & 0.3472 & 0.2060 & 0.1423 & 0.1074 & 0.0855 & 0.8420 & 0.4575 & 2.0682 \\
		& 2.0 & 0.3680 & 0.2310 & 0.1643 & 0.1260 & 0.0956 & 0.8401 & 0.3004 & 1.8488 \\
		\midrule
		2.0 & 0.5 & 0.1540 & 0.0610 & 0.0339 & 0.0223 & 0.0373 & 1.2537 & 1.8037 & 6.0888 \\
		& 1.0 & 0.2445 & 0.1087 & 0.0631 & 0.0425 & 0.0489 & 0.9046 & 1.1637 & 3.7830 \\
		& 1.5 & 0.3055 & 0.1472 & 0.0885 & 0.0607 & 0.0539 & 0.7602 & 0.8465 & 3.0349 \\
		& 2.0 & 0.3497 & 0.1792 & 0.1109 & 0.0772 & 0.0569 & 0.6822 & 0.6241 & 2.6781 \\
		\hline
	\end{tabular}
	\label{mo}
\end{table}

Table~\ref{mo} reports first four moments and shape characteristics of the TLXL distribution for various values of $\theta$ and $\alpha$. For fixed $\theta$, an increase in $\alpha$ generally raises the mean while reducing skewness and kurtosis, indicating a shift toward more symmetric and lighter-tailed behavior. Larger $\theta$ values are associated with a decrease in variance and coefficient of variation, reflecting a greater concentration of values. Overall, the distribution evolves from being highly skewed and leptokurtic to more symmetric and mesokurtic as the parameters increase.

\begin{figure}[H]
	\centering
	\subcaptionbox{}[3.2in]{\includegraphics[height=2.8in, width = 3.2in]{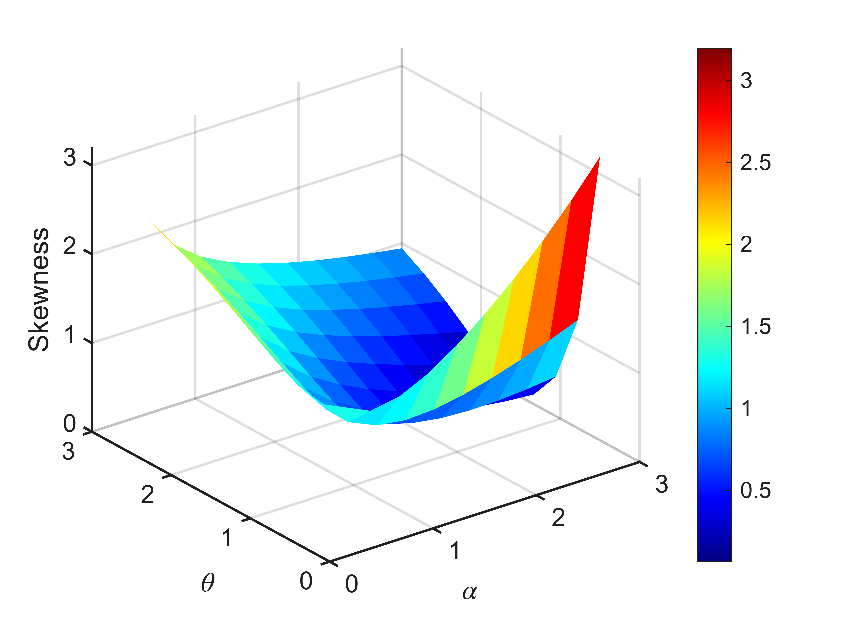}}
	\subcaptionbox{}[3.2in]{\includegraphics[height=2.8in, width = 3.2in]{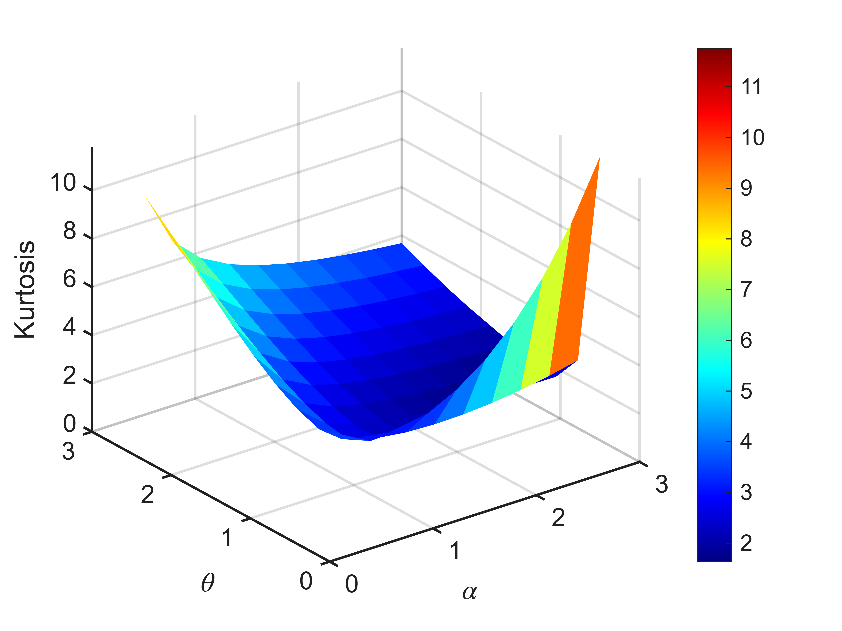}}
	\caption{$(a)$ Skewness surface of the TLXL distribution for different values of $\alpha$ and $\theta$, and $(b)$ Kurtosis surface of the TLXL distribution for various combinations of $\alpha$ and $\theta$.}
	\label{skewness}
\end{figure}

Figure \ref{skewness} illustrates the behavior of skewness and kurtosis surfaces of the TLXL distribution across different values of $\alpha$ and $\theta$.
It is observed that for small parameter values, the distribution remains nearly symmetric with moderate kurtosis, resembling the normal distribution.
However, as either $\alpha$ or $\theta$ increases, the skewness rises, indicating stronger asymmetry, while the kurtosis grows substantially, reflecting heavier tails and greater peakedness. This confirms the flexibility of the TLXL distribution in modeling both symmetric, (light-tailed dat)a and highly skewed (heavy-tailed data).

The moment generating function of the TLXL distribution (denoted by $M_X(t)$ where $t>0$)
can be defined as follows:
{\small
 \begin{eqnarray}
 	M_X(t) &=& E[\exp\{tX\}] = \int_0^{\infty} \exp\{t x\}  g_{TLXL}(x| \alpha,\theta)  dx \nonumber\\
  	&=& \frac{2\alpha \theta^2}{(1+\theta)^2} \sum_{l=0}^{\infty} \sum_{m=0}^{2l+1} \binom{\alpha-1}{l} \binom{2l+1}{m} \bigg(\frac{\theta}{(1+\theta)^2}\bigg)^m (-1)^l
 	\int_{0}^{1}\frac{ x^{m}(2+\theta+x) }{\exp\{(2\theta(l+1)-t)x\} } dx \nonumber\\
 	&=& \frac{2\alpha \theta^2}{(1+\theta)^2} \sum_{l=0}^{\infty} \sum_{m=0}^{2l+1} \sum_{n=0}^{\infty} \binom{\alpha-1}{l} \binom{2l+1}{m} \bigg(\frac{\theta}{(1+\theta)^2}\bigg)^m \frac{(-1)^{l+n}(2\theta(l+1)-t)^n}{n!} \nonumber\\
 	&~& \times \int_{0}^{1} t^{m+n} (t+2+\theta) dt \nonumber\\
 	&=& \frac{2\alpha \theta^2}{(1+\theta)^2} \sum_{l=0}^{\infty} \sum_{m=0}^{2l+1} \sum_{n=0}^{\infty} \binom{\alpha-1}{l} \binom{2l+1}{m} \bigg(\frac{\theta}{(1+\theta)^2}\bigg)^m \frac{(-1)^{l+n}(2\theta(l+1)-t)^n}{n!} \nonumber\\&~&\times ((2+\theta) \beta(m+n+1,1)+ \beta(m+n+2,1)). \nonumber\\
 \end{eqnarray}
}

Now, the quantile function of the TLXL distribution is derived. The quantile function is the inverse of the cumulative distribution function, and is particularly useful for describing the distributional behavior of data and for generating random samples in simulation studies.
The quantile function of the TLXL distribution is defined by the relation $ G_{TLXL}(x| \alpha,\theta)=u$ where $0 < u < 1$, and denoted by $Q_{TLXL}(u| \alpha,\theta) = G^{-1}_{TLXL}(u| \alpha,\theta)=x$ is the inverse of the CDF of the TLXL distribution. Thus, the quantile function of the TLXL distribution is derived as:
\begin{eqnarray}\label{eq2.9}
 \sqrt{1-u^{1/\alpha}}=\frac{(1+\theta)^2+\theta Q_{TLXL}(u| \alpha,\theta)}{(1+\theta)^2} \exp\{-\theta Q_{TLXL}(u| \alpha,\theta)\}.
\end{eqnarray}
To derive the quantile function of the TLXL distribution, it becomes necessary to solve Equation (\ref{eq2.9}) involving an exponential term of the form $x \exp\{x\}$. This type of expression cannot be inverted using elementary functions. Therefore, we employ the Lambert W function, denoted by $W(\cdot)$ (refer \cite{corless1996lambert}), which satisfies the identity $W(x) \exp\{W(x)\} = x.$ By rewriting the equation into the appropriate form, the Lambert W function allows us to isolate and solve for the desired variable, thereby obtaining an explicit expression for the quantile function as:
\begin{eqnarray}\label{lam}
	 Q_{TLXL}(u| \alpha,\theta)= -\bigg[\frac{(1+\theta)^2}{\theta}+\frac{1}{\theta}W_{-1} \bigg(-(1+\theta)^2 \exp\{-(1+\theta)^2\} \sqrt{1-u^{1/ \alpha}} \bigg)\bigg].
\end{eqnarray}
The negative branch of the Lambert $W$ function, denoted by $W_{-1}$. The functional expression for the quantile function given in (\ref{lam}) can be used to generate random variates from the TLXL distribution.

\subsection{R\'{e}nyi  Entropy}
The R\'{e}nyi entropy provides a generalized framework for quantifying the uncertainty associated with a probability distribution. For the TLXL distribution with the PDF $g_{TLXL}(x|\alpha, \theta)$, the R\'{e}nyi entropy (refer \cite{renyi1961measures}) of order $\gamma (\neq 1)>0 $, is defined as:
\begin{eqnarray}\label{reny}
H_{TLXL}^{\gamma} &=& \frac{1}{1 - \gamma} \log \left( \int_{0}^{\infty} g_{TLXL}(x|\alpha,\theta)^{\gamma} \, dx \right).
\end{eqnarray}
The entropy order $\gamma$  controls the sensitivity to the tail behavior of the distribution. When  $\gamma \to 1$, the R\'{e}nyi entropy converges to the Shannon entropy (refer \cite{shannon1948mathematical}
). Now, focusing on the integral component of the R\'{e}nyi entropy and substituting the PDF of the TLXL distribution given in \eqref{pdf} into it, we obtain
{\small
\begin{eqnarray}\label{renyint}
\int_0^\infty g_{TLXL}(x|\alpha,\theta)^{\gamma} dx&=&\bigg(\frac{2\alpha \theta^2}{(1+\theta)^2}\bigg)^\gamma \sum_{l=0}^{\infty} \sum_{m=0}^{2\gamma l+\gamma} \binom{\gamma(\alpha-1)}{l} \binom{2\gamma l+\gamma}{m}  \bigg(\frac{\theta}{(1+\theta)^2}\bigg)^l (-1)^m  \nonumber \\
&~&\times \int_{0}^{1}  (2+\theta+x)^\gamma x^m  \exp\{-2\theta(l+1)\gamma x\} dx \nonumber \\
&=& \bigg(\frac{2\alpha \theta^2}{(1+\theta)^2}\bigg)^\gamma \sum_{l=0}^{\infty} \sum_{m=0}^{2\gamma l+\gamma} \sum_{n=0}^{\infty} \binom{\gamma(\alpha-1)}{l} \binom{2\gamma l+\gamma}{m} \bigg(\frac{\theta}{(1+\theta)^2}\bigg)^l  \nonumber \\
&~& \times \frac{(-1)^{l+n} (2\theta(l+1)\gamma)^n}{m!}\int_{0}^{1}  (2+\theta+x)^\gamma x^{m+n} dx.
\end{eqnarray}
}

	According to Binomial theorem (i.e $(x+y)^n = \sum_{k=0}^{n} \binom{n}{k} x^k y^{n-k}$), the expansion of  $(x+2+\theta)^\gamma$ is expressed as	$ \sum_{k=0}^\gamma \binom{\gamma}{k} x^k (2+\theta)^{\gamma-k}.$
	Therefore, (\ref{renyint}) can be written as 
{\small
	\begin{eqnarray}\label{renycom}
	\int_0^\infty g_{TLXL}(x|\alpha,\theta)^{\gamma} dx &=& \bigg(\frac{2\alpha \theta^2}{(1+\theta)^2}\bigg)^\gamma \sum_{l=0}^{\infty} \sum_{m=0}^{2\gamma l+\gamma} \sum_{n=0}^{\infty} \sum_{k=0}^{\gamma} \binom{\gamma(\alpha-1)}{l} \binom{2\gamma l+\gamma}{m}  \bigg(\frac{\theta}{(1+\theta)^2}\bigg)^l  \nonumber \\
	&~& \times \binom{\gamma}{k} (2+\theta)^{\gamma-k} \frac{(-1)^{l+n} (2\theta(l+1)\gamma)^n}{m!} \int_{0}^{1}   x^{m+n+k} dx  \nonumber \\
	&=& \bigg(\frac{2\alpha \theta^2}{(1+\theta)^2}\bigg)^\gamma \sum_{l=0}^{\infty} \sum_{m=0}^{2\gamma l+\gamma} \sum_{n=0}^{\infty} \sum_{k=0}^{\gamma} \binom{\gamma(\alpha-1)}{l} \binom{2\gamma l+\gamma}{m}  \bigg(\frac{\theta}{(1+\theta)^2}\bigg)^l   \nonumber \\
&~& \times \binom{\gamma}{k} (2+\theta)^{\gamma-k} \frac{(-1)^{l+n} (2\theta(l+1)\gamma)^n}{m!}  \beta [m+n+k+1,1]. 
\end{eqnarray}
}
Substitute (\ref{renycom}) into (\ref{reny}), then we have
{\small
\begin{eqnarray}
	\nonumber H_\gamma(f) &=& \frac{1}{1 - \gamma} \log \biggl\{\bigg(\frac{2\alpha \theta^2}{(1+\theta)^2}\bigg)^\gamma \sum_{l=0}^{\infty} \sum_{m=0}^{2\gamma l+\gamma} \sum_{n=0}^{\infty} \sum_{k=0}^{\gamma} \binom{\gamma(\alpha-1)}{l} \binom{2\gamma l+\gamma}{m}  \bigg(\frac{\theta}{(1+\theta)^2}\bigg)^l \binom{\gamma}{k} \\
	&~& \times(2+\theta)^{\gamma-k}  \frac{(-1)^{l+n} (2\theta(l+1)\gamma)^n}{m!} \beta [m+n+k+1,1] \biggr\}.
\end{eqnarray}
}

\section{Parameter Estimation and Inference}
In this section, we present the maximum likelihood estimation method for estimating the parameters of the proposed TLXL distribution. Additionally, a Monte Carlo simulation study is carried out to examine the performance of the maximum likelihood estimates (MLEs) based on bias and mean squared error (MSE) across different sample sizes.
\subsection{Maximum Likelihood Estimation}
Let $(x_1,x_2,x_3,...,x_n)$ be a random sample of size $n$ drawn from the TLXL distribution with parameter vetor $\omega=(\theta,\alpha)$. Then, the maximum likelihood function (denoted by ${l(\omega)}$) is defined as follows:
\begin{eqnarray}
{l(\omega)}=\prod_{i=1}^{n}\bigg[ \frac{2\alpha\theta^2(2+x_i+\theta)\phi(\theta,x_i)\exp\{-2\theta x_i\}}{(1+\theta)^2}\left(1-\left(\phi(\theta, x_i)\exp\{-\theta x_i\}\right)^2\right)^{\alpha-1}\bigg].
\end{eqnarray}
The log-likelihood function, denoted by $\mathcal{L(\omega)}=\ln (l(\omega))$, is obtained as
\begin{eqnarray} \nonumber
	\mathcal{L(\omega)}&=&n \ln (2\alpha\theta^2)-2n\ln (1+\theta)+\sum_{i=1}^{n} \ln (2+\theta+x_i)-2\theta\sum_{i=1}^{n} x_i+ \sum_{i=1}^{n} \ln (\phi(\theta,x_i))\\&~&+(\alpha-1) \sum_{i=1}^{n}\ln (\eta(\theta, x_i))
\end{eqnarray}
where $\phi(\theta,x_i)=\frac{(1+\theta)^2+\theta x_{i}}{(1+\theta)^2}$ and $\eta(\theta, x_i)=1-\left(\phi(\theta, x_i)\exp\{-\theta x_i\}\right)^2$.
To obtain the MLEs of the TLXL distribution,
we compute the first-order partial derivatives of the log-likelihood function $\mathcal{L(\omega)}$ with respect to $\alpha$ and $\theta$ as given below:
\begin{eqnarray}
	\frac{\partial \mathcal{L(\omega)}}{\partial \alpha}&=& \frac{n}{\alpha} + \sum_{i=1}^{n} \ln \eta(\theta, x_i) \\
\mbox{and}~\frac{\partial \mathcal{L(\omega)}}{\partial \theta }&=&\frac{2n}{\theta(1+\theta)}+\sum_{i=1}^{n}\frac{1}{2+\theta+x_i}-2\sum_{i=1}^{n} x_i+\sum_{i=1}^{n} \frac{\phi^{'}(\theta, x_i)}{\phi(\theta, x_i)}+\sum_{i=1}^{n} \frac{\eta^{'}(\theta, x_i)}{\eta(\theta, x_i)}.\nonumber \\&~&
\end{eqnarray}
The MLEs, denoted as $\hat{\alpha}$ and $\hat{\theta}$, are obtained by solving the system of non-linear equations $\frac{\partial\mathcal{L(\omega)}}{\partial \alpha}=0$ and  $\frac{\partial\mathcal{L(\omega)}}{\partial \theta}=0$. Due to the nonlinear nature of these equations, a numerical technique such as the Newton–Raphson algorithm is required to obtain the parameter estimates.

\subsection{Simulation Study}
This section presents a simulation study aimed at evaluating the performance of the maximum likelihood estimators of the TLXL distribution parameters. The behavior of the estimators is examined in terms of bias and mean squared error (MSE) as the sample size increases. For each selected sample size, 1000 Monte Carlo replications are carried out. In each replication, the parameter estimates are obtained by minimizing the negative log-likelihood function through a numerical optimization procedure, which is implemented using the \texttt{R} programming environment with the L-BFGS-B optimization algorithm.
 Tables~\ref{table}, \ref{table1} and \ref{table3} present the average parameter estimates, along with their corresponding bias and MSE values. These results provide insights into the reliability and efficiency of the MLEs under different sample size conditions. The formulae for the bias and MSE are given as follows:
\begin{eqnarray*}
\text{Bias}(\hat{\Theta})=\frac{1}{N} \sum_{i=1}^{N} (\hat{\Theta}_i - \Theta) ~~\mbox{and}~~
\text{MSE}(\hat{\Theta})=\frac{1}{N} \sum_{i=1}^{N} (\hat{\Theta}_i - \Theta)^2,
\end{eqnarray*}
where $N$ is the number of replications, $\Theta$ is the true value of the parameter and $\hat{\Theta}_i$ is the estimated value of the parameter in the $i^{\text{th}}$ replication. The following steps are part of the research on simulations.
\begin{itemize}
\item[(i)] We consider the true values of the parameters as $(\theta, \alpha) = (0.5, 0.2), (1.5, 2)$ and $(2.5, 2)$ in Tables $2, 3$ and $4$, respectively.
\item[(ii)] In Table 2, the actual values of the reliability and hazard functions are $(\bar{G}(t|\theta,\alpha),\\ h(t|\theta,\alpha)) = (0.2414, 1.1240)$ and $(0.0961, 0.8244)$ for $t = (0.5, 1.5)$, respectively. At $t = 0.5, (1.5)$, the true values of the reliability function are $0.4814, (0.0407)$ and of the hazard function are $2.1530, (2.6196)$ in Table $3$, and the reliability function values are $0.1894, (0.0019)$ and the hazard function values are $4.3868, (4.6853)$ in Table $4$, respectively.
\item[(iii)] The various sample size values, $n = 10, 30, 50, 80,$ and $100$, are presented in the first column of Tables $2$, $3$ and $4$. The estimated values, bias, and MSE of the parameters $\theta$ and $\alpha$ are shown in the second and third columns, respectively. For $t = 0.5$, the fourth and fifth columns show the estimated values, biases, and MSEs of the reliability characteristics $\bar{G}(t|\theta,\alpha)$ and $h(t|\theta,\alpha)$. Similarly, the sixth and seventh columns correspond to the case $t = 1.5$.
\end{itemize}

\begin{table}[H]
	 \caption{ Average values, BIASs and MSEs of
		estimates for the parameters $\theta=0.5, \alpha=0.2 $, the reliability function $ 0.2414(0.0961)$ and hazard function $1.1240(0.8244)$ at the time $t=0.5(1.5)$.}
	\begin{center}
		\begin{tabular}{ccccccccccccc}
			\hline
			&   && $t=0.5$ &&&$t=1.5$&\\ \cline{4-5} \cline{7-8}\\
			
			&$\widehat{\theta}$ &$\widehat{\alpha}$& $\widehat{\bar{G}}(t|\theta,\alpha)$ & $\widehat{{h}}(t|\theta,\alpha)$ &&$\widehat{\bar{G}}(t|\theta,\alpha)$ & $\widehat{{h}}(t|\theta,\alpha)$\\
			
			&BIAS& BIAS& BIAS & BIAS && BIAS & BIAS\\
			
			$n$	&{MSE}&MSE& MSE & MSE &&MSE & MSE\\
			\hline

			10&0.7814 & 0.2523  &0.2392 &1.6208&& 0.0925  &1.3570  \\
			&0.2814 &0.0523 &-0.0022 &0.4968 && -0.0036  & 0.5327\\
			&0.8976 &0.0138 &0.0083& 3.2822&& 0.0032 &3.4755\\\\

			30&0.5930  &0.2113&0.2277& 1.2865&&0.0876& 0.9927 \\
			&0.0930 & 0.0113&-0.0138 &0.1625&&-0.0085& 0.1683\\
			&0.0796& 0.0022&0.0030&0.2489&&  0.0015&0.2622 \\\\
			
			50&0.5558  &0.2080& 0.2333&1.2191&& 0.0896&0.9204 \\
			&0.0558 &0.0080&-0.0082&0.0951&&-0.0065&0.0961\\
			&0.0236&  0.0009& 0.0020&0.0734&& 0.0010&0.0685\\\\
			
			80 &0.5530 &0.2076&0.2315&1.2146&& 0.0877 & 0.9143 \\
			&0.0530 &0.0076& -0.0099&0.0906&&-0.0085&0.0899\\
			&0.0163& 0.0009& 0.0013&0.0486&& 0.0008&0.0458\\\\
			
			100&0.5263  &0.2070 & 0.2394 &1.1664&& 0.0929&0.8679 \\
			&0.0263 & 0.0070& -0.0021&0.0424&&-0.0032& 0.0436 \\
			&0.0066& 0.0005&  0.0007&0.0192&&0.0004&0.0178\\
			\hline
		\end{tabular}
	\end{center}
	\label{table}
\end{table}

\begin{table}[H]
 \caption{ Average values, BIASs and MSEs of
		estimates for the parameters $\theta=1.5, \alpha=2 $, the reliability function $ 0.4814(0.0407)$ and hazard function $2.1530(2.6196)$ at the time $t=0.5(1.5)$.}
	\begin{center}
		\begin{tabular}{ccccccccccccc}
			\hline
			&   && $t=0.5$ &&&$t=1.5$&\\ \cline{4-5} \cline{7-8}\\
			
			&$\widehat{\theta}$ &$\widehat{\alpha}$& $\widehat{\bar{G}}(t|\theta,\alpha)$ & $\widehat{{h}}(t|\theta,\alpha)$ &&$\widehat{\bar{G}}(t|\theta,\alpha)$ & $\widehat{{h}}(t|\theta,\alpha)$\\
			
			&BIAS& BIAS& BIAS & BIAS && BIAS & BIAS\\
			
			$n$	&{MSE}&MSE& MSE & MSE &&MSE & MSE\\
			\hline
			10 &1.7721 & 3.2011  & 0.4877& 2.5236  && 0.0408 & 3.1764\\
			&0.2721& 2.2011 &0.0063&  	0.3706	  && 0.0002& 0.5568\\
			&0.4747& 1.5044 & 0.0163 &1.5426		   && 0.0011&1.9914 \\ \\
			
			30&1.5924& 2.2890&   0.4738 &2.3030		   &&0.0410 &  2.8117 \\
			&0.0924& 0.2890& -0.0077&0.1500		   &&0.0003 & 0.1922\\
			&0.1245& 0.9890& 0.0035&0.3243		    &&0.0006 &0.5238		     \\ \\
			
			50&1.5441& 2.1272&0.4793&2.2189	   		   && 0.0408 &2.7109\\
			&0.0441& 0.1272& -0.0021&0.0659		   &&0.0001&0.0913\\
			&0.0435& 0.2404&0.0026& 0.1211		    &&0.0003&0.1832\\\\
			
			80&1.5787& 2.1847&0.4741&2.2737	   		   && 0.0372 &2.7818\\
			&0.0787& 0.1847& -0.0073&0.1207		   &&-0.0035&0.1622\\
			&0.0409& 0.2619&0.0020& 0.1051		    &&0.0003&0.1731\\\\
			
			100&1.5437& 2.1148&0.4791&2.2154	   		   && 0.0390 &2.7096\\
			&0.0438& 0.1148& -0.0023&0.0624		   &&-0.0018&0.0900\\
			&0.0213& 0.1500&0.0013& 0.0532		    &&0.0001&0.0900\\
			
			\hline
		\end{tabular}
	\end{center}
	\label{table1}
\end{table}

\begin{table}[H]
	\caption{ Average values, BIASs and MSEs of
		estimates for the parameters $\theta=2.5, \alpha=2 $, the reliability function $ 0.1894(0.0019)$ and hazard function $4.3868(4.6853)$ at the time $t=0.5(1.5)$.}
	\begin{center}
		\begin{tabular}{ccccccccccccc}
			\hline
			&   && $t=0.5$ &&&$t=1.5$&\\ \cline{4-5} \cline{7-8}\\
			
			&$\widehat{\theta}$ &$\widehat{\alpha}$& $\widehat{\bar{G}}(t|\theta,\alpha)$ & $\widehat{{h}}(t|\theta,\alpha)$ &&$\widehat{\bar{G}}(t|\theta,\alpha)$ & $\widehat{{h}}(t|\theta,\alpha)$\\
			
			&BIAS& BIAS& BIAS & BIAS && BIAS & BIAS\\
			
			$n$	&{MSE}&MSE& MSE & MSE &&MSE & MSE\\
			\hline
			10	&2.9919   & 3.1858&   0.1825  & 5.3501 &&  0.0030 & 5.6848 \\
			&0.4919	  &1.1858 &-0.0069&0.9633 	&& 0.0012 & 0.9996 \\
			&1.5439   &	1.3651 &0.0070& 2.4115	&&0.0011& 2.3741\\ \\
			
			30	& 2.6679	&2.2851&  0.1817 &	4.7355 	   && 0.0027 & 5.0269 \\
			&0.1679		&0.2851& -0.0078 &0.3488 	   &&0.0008 & 0.3416\\
			& 0.3985	&0.9654&  0.0030&1.6276	       &&0.0009 &1.6564		     \\ \\
			
			50&2.5806&  2.1264 &0.1863&4.5493 	   		   && 0.0023 &4.8495\\
			& 0.0806& 0.1264&  -0.0031&0.1626		   &&0.0004&0.1642\\
			&0.1386&  0.2357&0.0017& 0.5609	    && 0.0008&0.5782\\\\
			
			80&2.6408& 2.1820&0.1790&4.6703   		   && 0.0018 &4.9726 \\
			&0.1408& 0.1820& -0.0104& 0.2835 	   &&-0.0001&0.2873\\
			&0.1301& 0.2561&0.0014& 0.5243	    &&0.0006&0.5427\\\\
			
			100&2.5788 & 2.1135&0.1844 & 4.5426   		   && 0.0019 &4.8460\\
			&0.0788& 0.1135& -0.0050&0.1558 	   && 0.0000&0.1607\\
			&0.0674&  0.1464&0.0008& 0.2679	    &&0.0003&0.28134\\
			\hline
		\end{tabular}
	\end{center}
	\label{table3}
\end{table}

The tabulated values reveal the following observations.
\begin{itemize}
	\item[(i)] The BIAS and MSEs of the proposed estimates tend to decrease as the sample size $n$ increases.
	\item[(ii)] The estimated values of the MLEs for all of the samples are becoming closer to the actual values of a and b as the sample size $n$ increases.
	\item[(iii)] The reliability characteristics, such as the hazard and reliability functions, perform better in terms of MSE at time $t=1.5$ than at time $t=0.5$.
\end{itemize}

\section{Application to Real Lifetime Data}
In this section, we analyze two real lifetime data sets for illustrative purpose. The data sets are consist of customers waiting times (in minutes) before receiving service at two different banks. These observations were originally reported by \cite{ghitany2008lindley} and were later further examined in \cite{ghitany2013power}. We revisit this data set here to evaluate the performance of the proposed model and to illustrate its applicability to real-world waiting-time scenarios. The competing models considered for comparison with the TLXL distribution are presented in Table \ref{table2}.

\begin{table}[H]
	 \centering  \caption{\label{table2}  The distribution models taken for the comparison with the TLXL distribution. }
	\begin{center}
		\renewcommand{\arraystretch}{1.5}
		\small
		\begin{tabular}{|p{2.3cm}|p{3.5cm}|p{8.0cm}|}
			\hline
			Distribution	&Author \& Year							& PDF \\
			\hline
			Weibull &	\cite{xie1996reliability}& $ f(x| k, \lambda)= \frac{k}{\lambda} \left(\frac{x}{\lambda}\right)^{k-1} e^{ \left(-\frac{x}{\lambda}\right)^k}; \text{ $ x>0, k,\lambda>0$}.$ \\
			Gamma & \cite{aksoy2000use} &
			$ f(x| \alpha, \beta) = \frac{\beta^\alpha}{\Gamma(\alpha)} x^{\alpha - 1} e^{-\beta x}; \text{   $x > 0, \alpha, \beta >0$.}$  \\
			Lindley & \cite{ghitany2008lindley}		 	& 	$
			f(x|\theta)=\frac{\theta^2}{\theta+1}(1+x)e^{-\theta x};\text{ $ x>0,\theta>0$}.	$   \\
			Exponential	&\cite{balakrishnan1996exponential}	  & $
			f(x| \lambda)=
			\lambda e^{-\lambda x};  \text{ $x > 0$, $\lambda>0$}.$\\
			\hline
		\end{tabular}
		\normalsize
	\end{center}
\end{table}
To facilitate the comparison of lifetime distributions, the values of $-\ln L$, the Akaike Information Criterion ($AIC=-2 \ln L + 2k$),  Akaike Information Criterion Corrected ($AICc=\mathrm{AIC} + \frac{2k(k+1)}{n - k - 1}$), Bayesian Information Criterion ($BIC=-2 \ln L + k \ln(n)$), Hannan–Quinn Information Criterion ($HQIC=-2 \ln L + 2k \ln(\ln(n))$), Kolmogorov-Smirnov Statistic (KS Stat.) and the p-value for the given data set have been calculated and are shown in Table \ref{goodness of fit 1} and \ref{goodness of fit 2}. The best distribution is the distribution which corresponds to lower values of  $-\ln L$, $AIC$, $AICc$, $BIC$ and K-S statistic and higher p-value.

\subsection{Data Set-1: Customers Waiting Times in Bank A }
The following data set represents the waiting times (in minutes) for 100 customers of the bank prior to receiving service in Bank~A. The data set is presented in Table~\ref{bank a}.
\begin{table}[H]
		\centering
		\caption{Observed customer waiting times (in minutes) at Bank~A.}
		\label{bank a}
		\begin{tabular}{p{15cm}}
			\hline
			0.8, 0.8, 1.3, 1.5, 1.8, 1.9, 1.9, 2.1, 2.6, 2.7, 2.9, 3.1, 3.2, 3.3, 3.5, 3.6, 4.0, 4.1, 4.2, 4.2, 4.3, 4.3, 4.4, 4.4, 4.6, 4.7, 4.7, 4.8, 4.9, 4.9, 5.0, 5.3, 5.5, 5.7, 5.7, 6.1, 6.2, 6.2, 6.2, 6.3, 6.7, 6.9, 7.1, 7.1, 7.1, 7.1, 7.4, 7.6, 7.7, 8.0, 8.2, 8.6, 8.6, 8.6, 8.8, 8.8, 8.9, 8.9, 9.5, 9.6, 9.7, 9.8, 10.7, 10.9, 11.0, 11.0, 11.1, 11.2, 11.2, 11.5, 11.9, 12.4, 12.5, 12.9, 13.0, 13.1, 13.3, 13.6, 13.7, 13.9, 14.1, 15.4, 15.4, 17.3, 17.3, 18.1, 18.2, 18.4, 18.9, 19.0, 19.9, 20.6, 21.3, 21.4, 21.9, 23.0, 27.0, 31.6, 33.1, 38.5 \\
			\hline
		\end{tabular}
	\end{table}

The descriptive statistics of the 100 waiting-time observations shows substantial variability across customers. The average waiting time is about 9.88 minutes, whereas the median value of 8.1 minutes suggests a right-skewed distribution influenced by several large observations, with the longest wait reaching 38.5 minutes. The minimum wait is 0.8 minutes, producing a broad range of 37.7 minutes. The data exhibit considerable dispersion, as indicated by a standard deviation of approximately 7.24 minutes. Moreover, the respective first and third quartiles, 4.675 and 13.025 minutes, produce an interquartile range of 8.35 minutes, further highlighting the spread and skewness present in the distribution.

In the following, Table \ref{Esti1} presents the estimated parameter values for the competing distribution models on Data Set-1.
\begin{table}[H]
	 \centering \caption{Estimated parameter values for Data Set-1 under different distributions.}
	\begin{center}
		\renewcommand{\arraystretch}{1.5}
		\begin{tabular}{|c|c|}
			\hline
			Distribution(s)	& Est.Parameter values of Data Set-1\\
			\hline
			TLXL  	 & $\hat{\theta}=0.1104$, $\hat{\alpha}=1.1002$  \\
			
			Weibull  & $\hat{k}=1.4585$, $\hat{\lambda}=10.9553$ \\
			
			Gamma 		&	 $ \hat{\alpha}=2.0088$, $\hat{\beta}=0.2034$\\
			Lindley	 	& 	 $\hat{\theta}=0.1866$ \\
			Exponential	& $\hat{\lambda}=0.1012$ \\
			\hline
		\end{tabular}
	\end{center}
		 \label{Esti1}
\end{table}

\begin{figure}[H]
		\centerline{
		\begin{tabular}{cc}
\subcaptionbox{\label{c1}}[3in]{\includegraphics[height=2in, width = 3in]{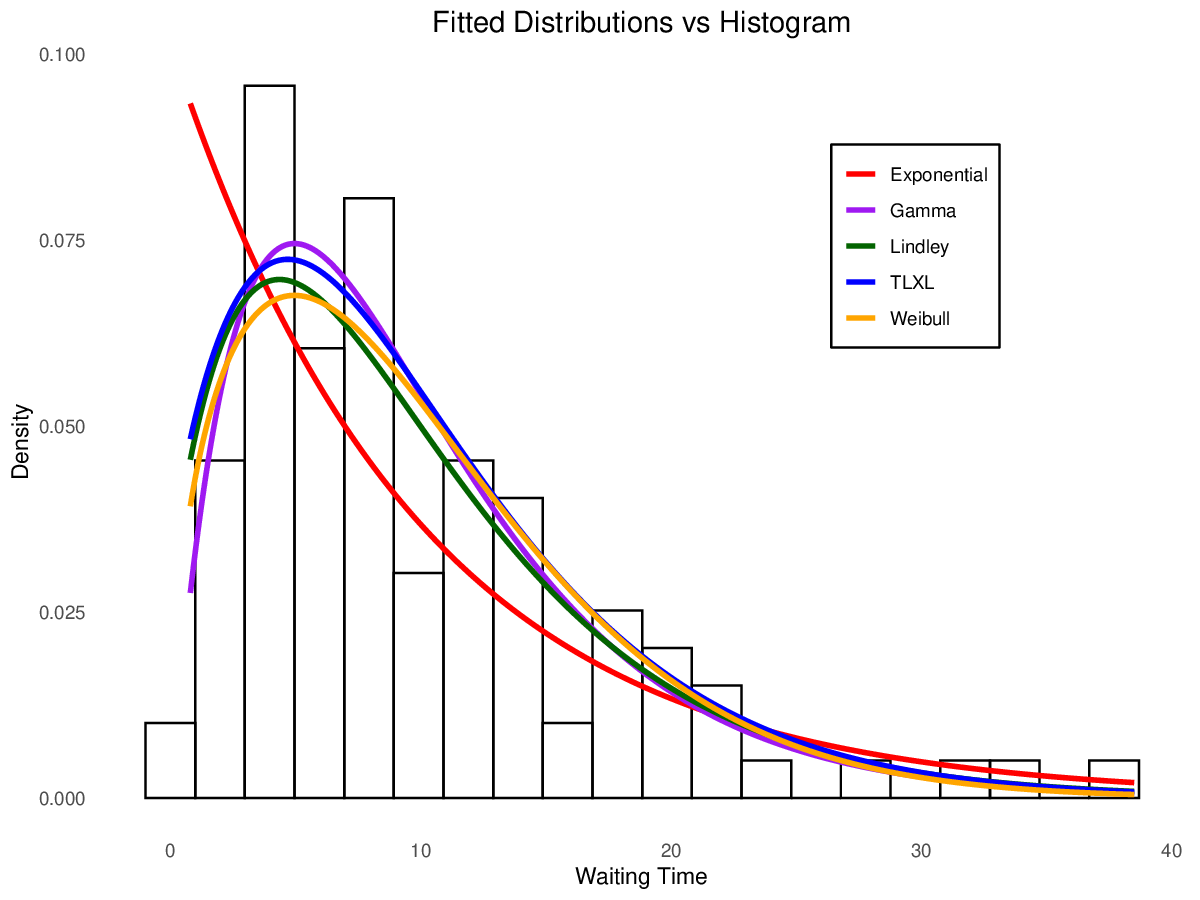}}
			\subcaptionbox{\label{c2}}[3in]{\includegraphics[height=2in, width = 3in]{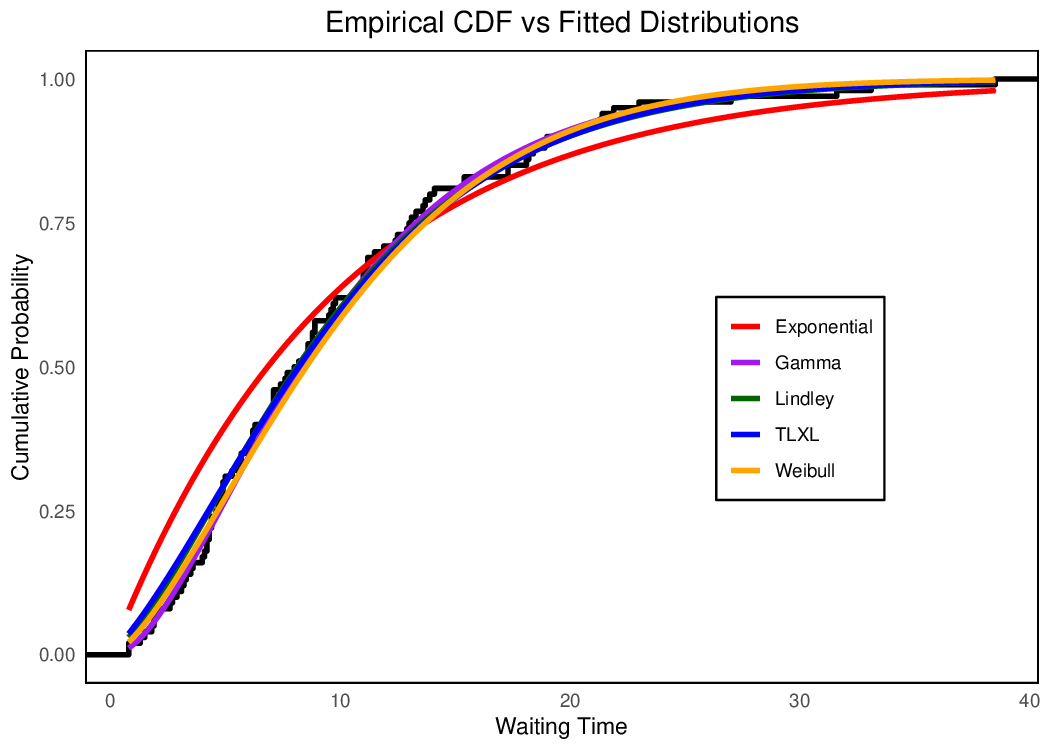}}\\
			\subcaptionbox{\label{c3}}[3in]{\includegraphics[height=2in, width = 3in]{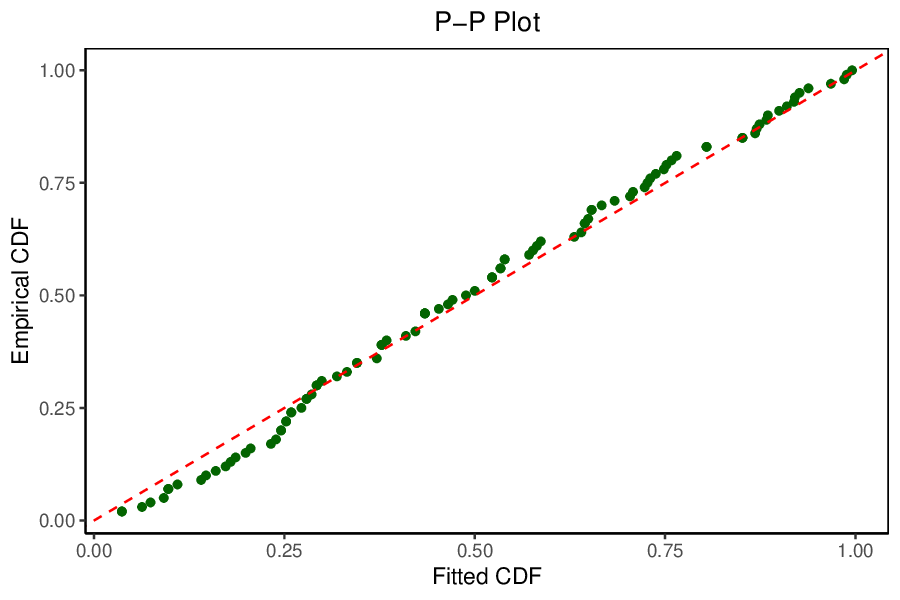}}
			\subcaptionbox{\label{c4}}[3in]{\includegraphics[height=2in, width = 3in]{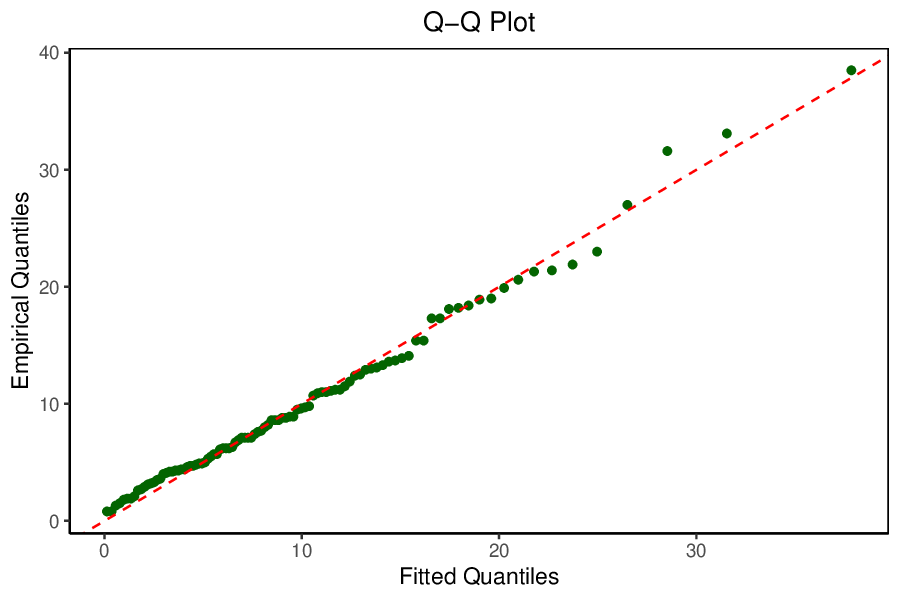}}\\
			\subcaptionbox{\label{c5}}[3in]{\includegraphics[height=2in, width = 3in]{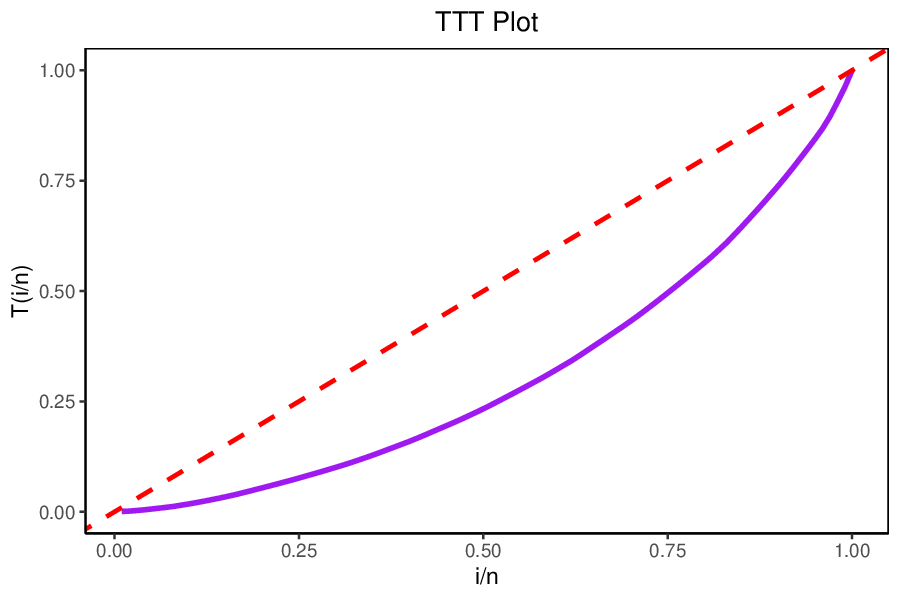}}
			\subcaptionbox{\label{c6}}[3in]{\includegraphics[height=2in, width = 3in]{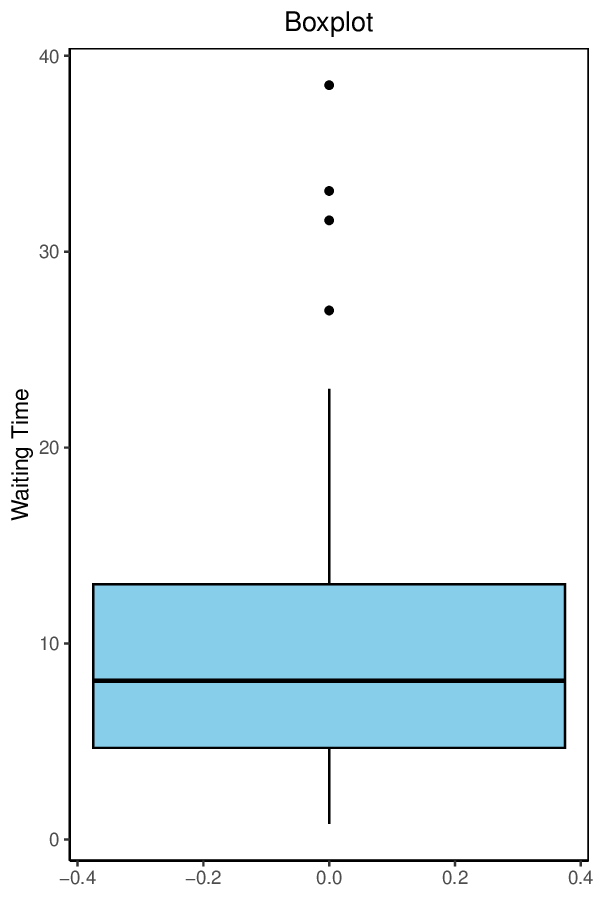}}
	\end{tabular}}
	\caption{$(a)$ The histograms along with the probability density function plots of the fitted Exponential, Gamma, Lindley and Weibull models for the real data sets are shown; (b) shows the ECDF; (c) presents the P–P plots; (d) displays the Q–Q plots; (e) TTT plot; and (f) box plot for the bank customer waiting time data in Bank A.}
\end{figure}

\begin{table}[H]
	\centering
	\caption{Goodness-of-fit assessment of different distributional models for Data Set-1 based on negative log-likelihood and information-criterion metrics}
	\small
	\setlength{\tabcolsep}{4pt}
	\begin{tabular}{l r r r r r r r}
		\hline
		\textbf{Dist.}  & \textbf{$-\ln L$} & \textbf{AIC} & \textbf{AICc} & \textbf{BIC} & \textbf{HQIC} & \textbf{KS Stat.} & \textbf{ $p$-value} \\
		\hline
		TLXL      & 312.5764 & 629.1528 & 629.2765 & 634.3631 & 631.2615 & 0.0325 & 0.9946 \\
		Weibull        & 318.7307 & 641.4614 & 641.5851 & 646.6717 & 643.5701 & 0.0578 & 0.8921 \\
		Gamma         & 317.3001 & 638.6002 & 638.7240 & 643.8106 & 640.7090 & 0.0425 & 0.8736 \\
		Lindley         & 319.0374 & 640.0748 & 640.1156 & 642.6800 & 641.1292 & 0.0677 & 0.7494 \\
		Exponential         & 329.0209 & 660.0418 & 660.0826 & 662.6469 & 661.0961 & 0.1730 & 0.0050 \\
		\hline
	\end{tabular}
	\normalsize
	\label{goodness of fit 1}
\end{table}
Table~\ref{goodness of fit 1}, shows the evaluation metrics for five fitted distributions. Among them, the TLXL distribution provided the best fit to the data, evidenced by the lowest values across all information criteria and the highest KS test p-value (0.9946). Figure~\ref{c1} displays the histogram and fitted probability density functions for the customer waiting times data set. The density curves were estimated using the maximum likelihood method. It can be observed that the TLXL distribution offers the best fit, effectively capturing the majority of the data compared to the competing models. Figure~\ref{c2} presents the empirical cumulative distribution function (ECDF) alongside the fitted cumulative distribution functions for the observed data. Further, Figures~\ref{c3} and~\ref{c4} illustrate the probability-–probability (P–P) plot and the quantile–-quantile (Q–Q) plot for the TLXL distribution, respectively,  both of which show that the empirical points closely follow the diagonal line, indicating a good agreement between the theoretical and observed quantiles.  Additionally, Figure~\ref{c5} displays the Total Time on Test (TTT) plot, which suggests a decreasing hazard rate consistent with the fitted distribution, while Figure~\ref{c6} presents the box plot, highlighting the right-skewed nature of the data and the presence of a few extreme observations.

Based on the visual assessments from the histogram, ECDF, P–P, Q–Q, TTT and box plots, the TLXL distribution provides the best overall fit to the empirical data. It closely follows the ECDF and density with minimal deviation across the entire range. Compared to other models, TLXL distribution more accurately captures both the central body and the tail behavior, making it the most appropriate distribution for this data set.

\subsection{Data Set-2: Customers Waiting Times in Bank B}
The following data set represents the waiting times (in minutes) for 60 customers of the bank prior to receiving service in Bank~B.
	\begin{table}[H]
	\centering
	\caption{Observed customer waiting times (in minutes) at Bank~B.}
	\label{bank b}
	\begin{tabular}{p{15cm}}
		\hline
		0.1, 0.2, 0.3,0.7, 0.9, 1.1, 1.2, 1.8, 1.9, 2.0, 2.2, 2.3, 2.3, 2.3, 2.5, 2.6, 2.7, 2.7,2.9, 3.1, 3.1, 3.2, 3.4, 3.4, 3.5, 3.9, 4.0, 4.2, 4.5, 4.7, 5.3, 5.6, 5.6,6.2, 6.3, 6.6, 6.8, 7.3, 7.5, 7.7, 7.7, 8.0, 8.0, 8.5, 8.5, 8.7, 9.5,10.7, 10.9, 11.0, 12.1, 12.3, 12.8, 12.9, 13.2, 13.7, 14.5, 16.0,16.5, 28.0 \\
		\hline
	\end{tabular}
\end{table}
The descriptive analysis of the 60 customers waiting time observations shows that there is variability in the data set. The mean waiting time is approximately 6.37 minutes, while the median is 5 minutes. This indicates that the data is right-skewed by the presence of higher values such as the maximum of 28 minutes. The minimum waiting time is 0.1 minutes, while the range is 27.9 minutes. Additionally, the standard deviation is about 5.19 minutes, indicating moderate variability in the data set. Moreover, the first and third quartiles of the data set are 2.575 and 8.55 minutes, respectively. This indicates that the interquartile range is 5.98 minutes.

Table \ref{Esti2} presents the estimated parameter values for the competing distribution models on Data Set-2.

\begin{table}[H]
	\centering
	 \caption{Estimated parameter values for Data Set-2 under different distributions.} 

	\begin{center}
		\renewcommand{\arraystretch}{1.5}
		\begin{tabular}{|c|c|}
			\hline
			Distribution(s)	& Est.Parameter value of Data Set-2\\
			\hline
			TLXL  	 &$\hat{\theta}=0.1614$, $\hat{\alpha}=1.2012$ \\
			
			Weibull  &$\hat{k}=1.2351$, $\hat{\lambda}=6.8115$ \\
			
			Gamma 		&$\hat{\alpha}=1.3764$, $\hat{\beta}=1.4585$\\
			Lindley	 	 &$\hat{\theta}=0.2797$\\
			Exponential	 &$\hat{\lambda}=0.1570$\\
			\hline
		\end{tabular}
	\end{center}
		 \label{Esti2}
\end{table}

\begin{table}[H]
	\centering
	\caption{Goodness-of-fit assessment of different distributional models for Data Set-2 based on negative log-likelihood and information-criterion metrics}
	\small
	\setlength{\tabcolsep}{4pt}
	\begin{tabular}{l r r r r r r r}
		\hline
		\textbf{Dist.}  & \textbf{$-\ln L$} & \textbf{AIC} & \textbf{AICc} & \textbf{BIC} & \textbf{HQIC} & \textbf{KS Stat.} & \textbf{ $p$-value} \\
		\hline
		TLXL      & 165.4307 & 334.8614 & 335.0719 & 339.0501 & 336.4998 & 0.0611 & 0.9786 \\
		Weibull  & 169.1168 & 342.2335 & 342.4441 & 346.4222 & 343.8720 & 0.0998 & 0.5887 \\
		Gamma         & 169.3625 & 342.7250 & 342.9355 & 346.9136 & 344.3634 & 0.0722 & 0.9742 \\
		Lindley        & 169.1014 & 340.2029 & 340.2719 & 342.2972 & 341.0221 & 0.0797 & 0.8401 \\
		Exponential  & 171.0803 & 344.1605 & 344.2295 & 346.2549 & 344.9797 & 0.1295 & 0.2663 \\
		\hline
	\end{tabular}
	\normalsize
	\label{goodness of fit 2}
\end{table}
As observed in Table \ref{goodness of fit 2}, it is clear that the TLXL model has the highest performance, as it has recorded the lowest values for all information criteria (AIC, AICc, BIC, and HQIC), as well as the lowest KS statistic and highest p-value. The results clearly prove that it is the best model in terms of fit. The Gamma distribution is observed to be in close competition with similar fit measures and an equally high p-value. The performance of the Weibull and Lindley models is also moderate. On the contrary, it is observed in Table \ref{goodness of fit 2} that the exponential model has recorded the worst performance, as it has recorded high information criteria values and has recorded the highest KS statistic.

\begin{figure}[H]
	\centerline{
		\begin{tabular}{cc}
			\subcaptionbox{}[3in]{\includegraphics[height=2in, width = 3in]{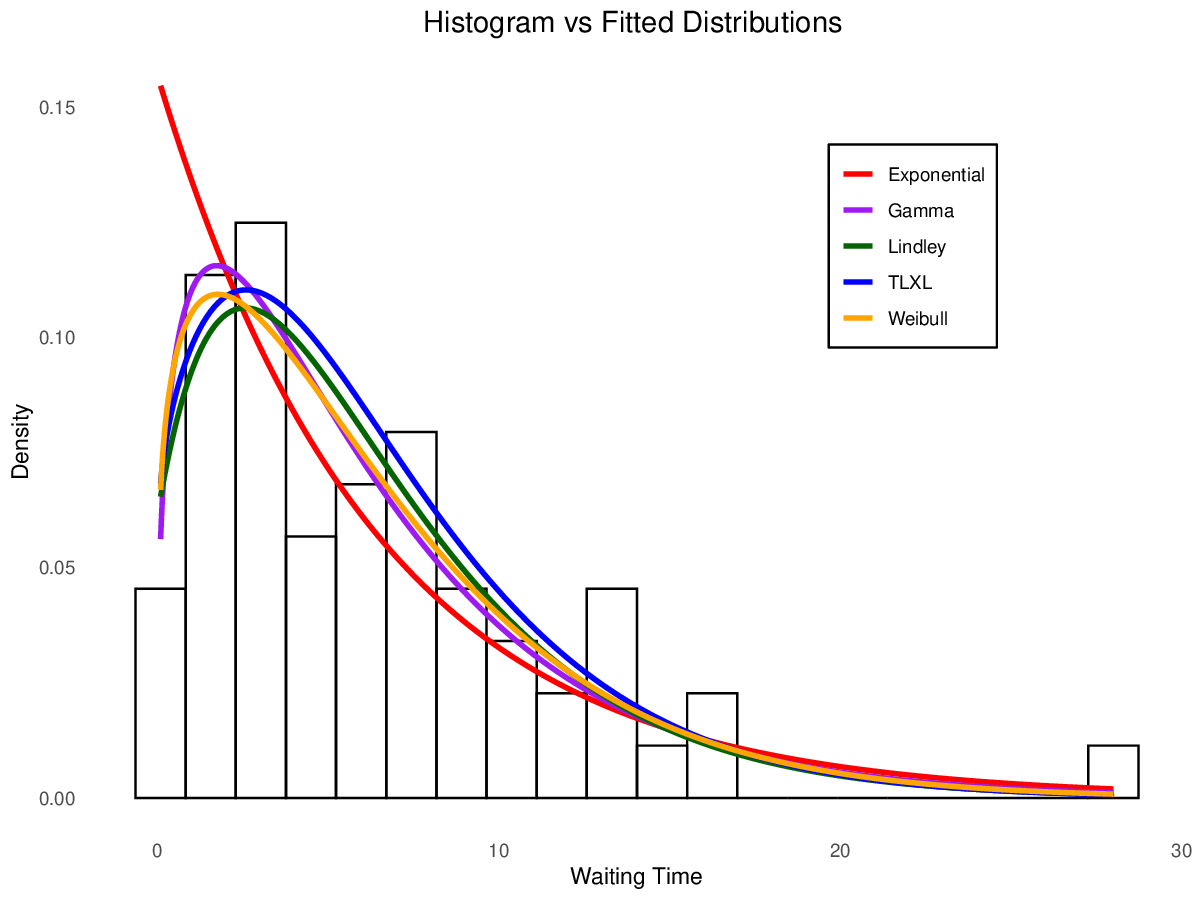}}
			\subcaptionbox{}[3in]{\includegraphics[height=2in, width = 3in]{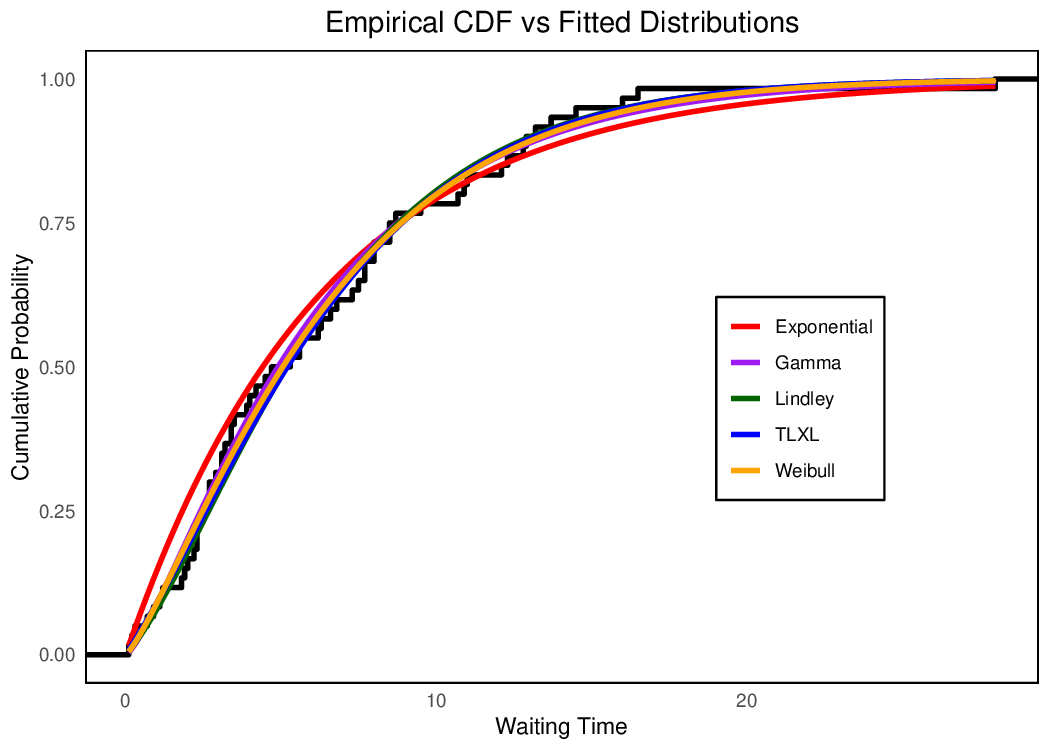}}\\
			\subcaptionbox{}[3in]{\includegraphics[height=2in, width = 3in]{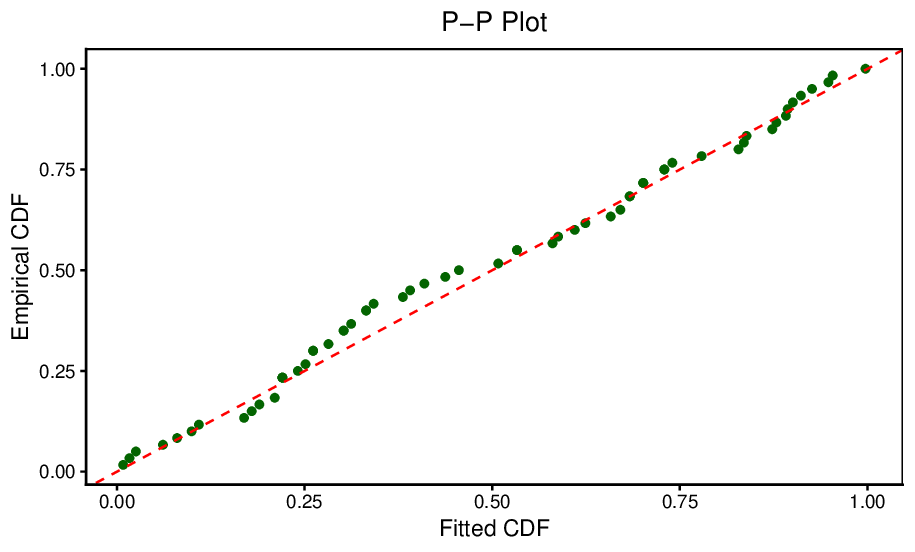}}
			\subcaptionbox{}[3in]{\includegraphics[height=2in, width = 3in]{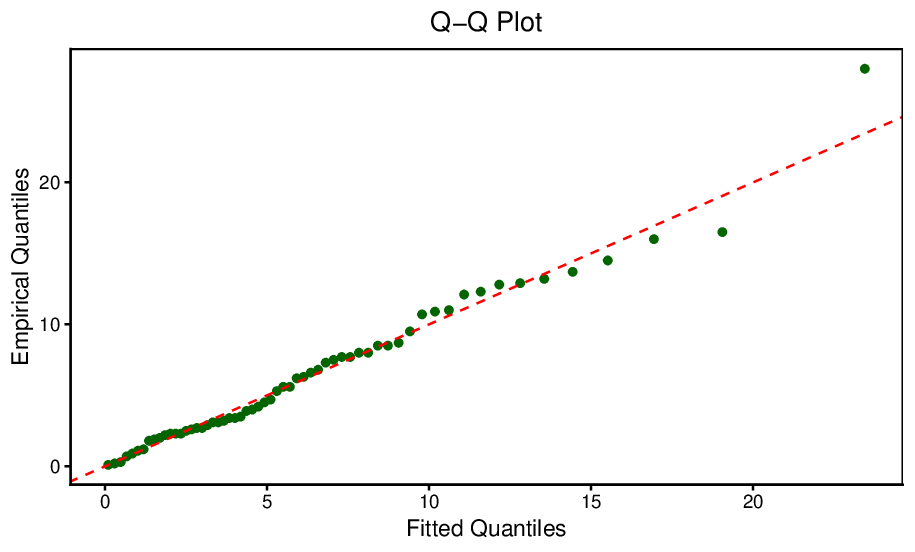}}\\
			\subcaptionbox{}[3in]{\includegraphics[height=2in, width = 3in]{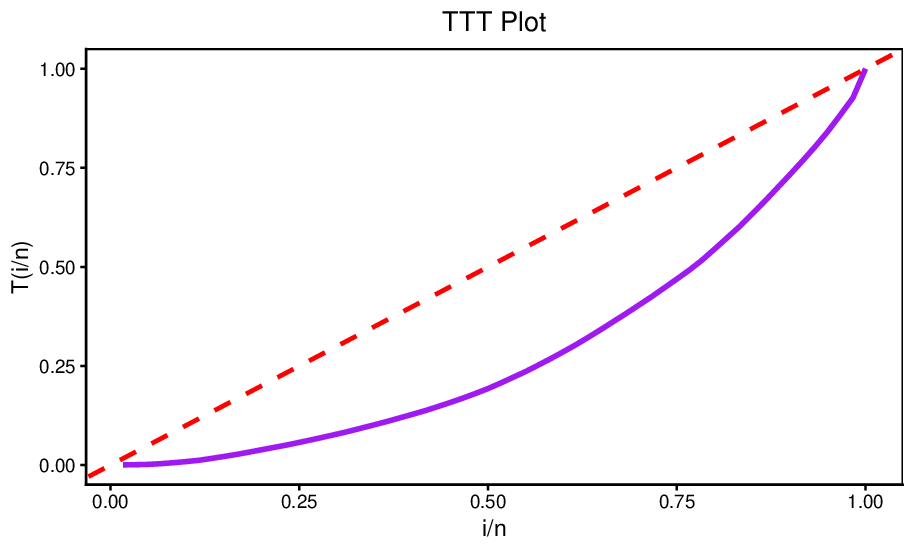}}
			\subcaptionbox{}[3in]{\includegraphics[height=2in, width = 3in]{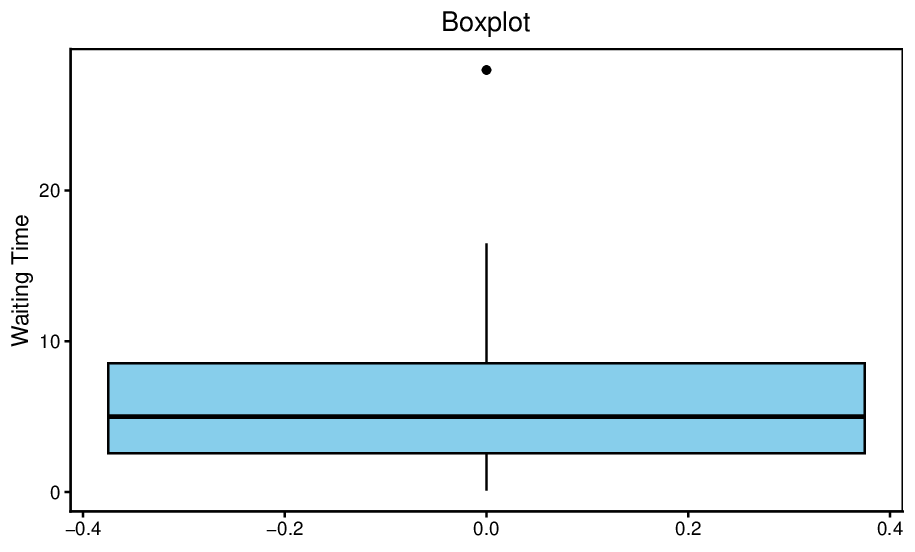}}
	\end{tabular}}
	\caption{$(a)$ The histograms along with the probability density function plots of the fitted Exponential, Gamma, Lindley and Weibull models for the real data sets are shown; (b) shows the ECDF; (c) presents the P–P plots; (d) displays the Q–Q plots; (e) TTT plot; and (f) box plot for the bank customer waiting time data in Bank B.}
	\label{fitted2}
\end{figure}

Figure~\ref{fitted2}, provides several visual assessments of the fitted models, and it is evident that the TLXL distribution aligns most closely with the observed waiting time data. In both the histogram and ECDF plots, the TLXL curve follows the empirical pattern more accurately than the other models. The P--P and Q--Q plots also show points lying nearer to the reference line for TLXL, confirming its superior fit. Overall, the visualizations clearly indicate that TLXL is the best-fitting distribution.

\section{Concluding Remarks and Future Directions}
In the present paper, a new model of lifetime distribution called as the Topp–Leone X-Lindley distribution has been proposed, and the major statistical and reliability properties of the proposed distribution have been studied. To estimate the parameters of the TLXL distribution, the maximum likelihood estimation has been employed. A detailed simulation study has been carried out to investigate the performance of the estimators in terms of bias and MSE. To show the applicability of the proposed distribution in practice, the ToppLeoneX-Lindley distribution has been applied to some real lifetime data sets, and the distribution has been shown to perform better than some other well-known distributions such as Exponential, Lindley, Gamma, and Weibull. Future studies may also investigate the Bayesian estimation procedures for the TLXL distribution, as well as the extension of the TLXL distribution in the context of regression and censoring models. Additionally, further generalizations of the TLXL distribution by employing different families of generators may be considered. 

\section*{Declarations}
\subsection*{Ethics Approval}
Not applicable. This study did not involve human participants, human data, or animal subjects. It is a theoretical/methodological contribution based on statistical distribution modeling and does not require ethics committee approval.

\subsection*{Funding}
This research received no specific grant from any funding agency.

\subsection*{Competing Interests}
The authors declare no competing interests.

\subsection*{Availability of Data}
All data used in this study are presented within the manuscript, and their original sources have been appropriately cited.

\bibliography{references} 
\end{document}